\documentstyle[prl,aps,epsf,multicol]{revtex}
\begin{document}
\draft
\tighten

\title{Dirt Softens Soap: Anomalous Elasticity of Disordered Smectics}
\author{Leo Radzihovsky} 
\address{Department of Physics, University of Colorado,
Boulder, CO 80309}
\author{John Toner} \address{Dept. of Physics,
Materials Science Inst., and Inst. of Theoretical Science,
University of Oregon, Eugene, OR 97403}

\date{\today}
\maketitle
\begin{abstract}

We show that a smectic in a disordered medium (e.g., aerogel) exhibits
anomalous elasticity, with the compression modulus $B({\bf k})$
vanishing and the bend modulus $K({\bf k})$ diverging as ${\bf k}
\rightarrow {\bf 0}$. In addition, the effective disorder develops
long ranged correlations. These divergences are {\it much} stronger
than those driven by thermal fluctuations in pure smectics, and are
controlled by a zero temperature glassy fixed point, which we study in
an $\epsilon=5-d$ expansion. We discuss the experimental implications
of these theoretical predictions.

\end{abstract}

%\twocolumn
\begin{multicols}{2}
\narrowtext

The effects of quenched disorder on the properties of condensed matter
systems continues to be a fascinating area of active research, which
includes the study of disordered superconductors\cite{VG}, charge
density waves\cite{cdw}, Josephson junction arrays\cite{JJ}, and
Helium in aerogel\cite{HeAerogel}, to name a few. Some of this
attention has focussed\cite{Clark,RTaerogel} on liquid crystals in the
random environment of an aerogel. While a complete picture of
aerogel-confined liquid crystals is still being
developed\cite{RTaerogel}, in this Letter we show that a smectic phase
of these systems possesses strong anomalous elasticity, when subjected
to an arbitrarily weak amount of quenched disorder. This result has
important experimental consequences.

The anomalous elasticity of {\em pure} smectics was predicted sometime
ago\cite{GP}, and is characterized by bulk compressional and tilt
moduli $B({\bf k})$ and $K({\bf k})$ which, respectively, vanish and
diverge at long wavelengths (${\bf k}\rightarrow {\bf 0}$).  This is a
general property of all one-dimensional crystals in which the
direction of the 1d ordering wavevector is chosen spontaneously. As a
consequence of this {\em spontaneous} breaking of rotational symmetry,
(a property of smectics but not of charge density waves) in such
systems, compression can be relieved by smoothing out fluctuations
(wrinkles in the smectic layers), so the effective layer compressional
modulus $B({\bf k})$ vanishes at long wavelengths.  Similarly, in the
presence of fluctuations, a bending of smectic layers necessarily
leads to a compression, which implies that the effective tilt modulus
$K({\bf k})$ diverges at long wavelengths.  Unfortunately, this {\em
thermally} driven behavior in {\em pure} smectics is difficult to
observe experimentally, because the effect is very weak (logarithmic)
in 3d.

The main ingredients necessary for anomalous elasticity, namely {\em
spontaneously} broken rotational invariance, and fluctuations, both
exist even at zero temperature in quenched disordered smectics. In
this Letter we demonstrate the existence of anomalous elasticity in
quenched-disordered smectics which is significantly stronger than the
marginal anomalous elasticity of thermal smectics, and is described by
a zero-temperature fixed point that is perturbatively accessible in
$d=5-\epsilon$ dimensions.  The elastic anomaly is much stronger in
quenched disordered smectics because layer fluctuations are much
larger, even at $T=0$, than in a pure smectic at $T>0$.

One experimental signature of these divergences is in the smectic
correlation length for smectics in aerogel, which has a different,
universal dependence on the bare smectic elastic constants and the
aerogel density than predicted by harmonic theory.  Detailed
predictions for these lengths can be found at the end of this Letter.

Our model of a quenched disordered smectic starts with de Gennes'
theory\cite{deGennes} for the smectic density field $\psi$ and the
nematic director $\hat{\bf n}$, and includes a disorder field that
couples to the nematic director, via
\begin{eqnarray}
\delta H_{d}= - \int d^d r \big({\bf g}({\bf r})\cdot{\bf n}\big)^2\;,
\label{Fn}
\end{eqnarray}
where $\bf g(\bf r)$ is a quenched random vector along the locally
preferred nematic alignment. Such preferences can arise because, e.g.,
the nematogens may align with the local, randomly oriented aerogel
strands.  Below the nematic--smectic-A transition $\psi$ can be
written as $|\psi|e^{i q_0 u({\bf r})}$, with $u({\bf r})$ describing
the local displacement of the smectic layers from perfect
periodic order. Furthermore, far below the nematic transition
(i.e. inside the smectic phase) we can take $\hat{\bf n}\approx
\hat{\bf z}+\delta{\bf n}$, where $\hat{\bf z}$ is the mean normal to
the smectic layers, and $\perp$ denotes directions orthogonal to
$\hat{\bf z}$. Integrating out the ``massive'' $|\psi|$ and
$\delta{\bf n}$ fields, which has the effect of replacing $\delta{\bf
n}\rightarrow \bbox{\nabla}_\perp u$ (i.e., the Higg's mechanism), and
keeping only the most relevant terms, we obtain the elastic
Hamiltonian that defines our model,
\begin{eqnarray}
\hspace{-.3in}H&=&\int_{\bbox r}\bigg[{K\over 2}(\nabla^2_\perp {u})^2
+ {B\over 2}\big(\partial_z u - 
{1\over2}(\bbox{\nabla}_\perp u)^2\big)^2
+{\bf h}({\bf r})\cdot\bbox{\nabla_\perp} u\bigg]\;,\label{H}\nonumber\\
\end{eqnarray}
%
%\nonumber\\
%&+&{\bf h}({\bf r})\cdot\bbox{\nabla_\perp} u\bigg]\;,
%\label{H}
%\end{eqnarray}
%
where ${\bf h}({\bf r})\equiv g_z({\bf r}){\bf g}({\bf r})$ is
quenched random tilt disorder that for simplicity we take to be
Gaussian, zero-mean, and completely characterized by
\begin{equation}
\overline{h_i({\bf r})h_j({\bf r'})}=
\Delta\delta^d({\bf r}-{\bf r'})\delta_{ij}\;.
\label{Delta}
\end{equation}
The use of short-ranged correlations, even though the {\em density} of
the (fractal) aerogel has long-ranged correlations, is justified,
since the {\em orientations} of its constituent silica strands, beyond
their microscopic persistance length, are certainly short-range
correlated\cite{RTaerogel,Clark}. We will focus here on the bahavior
of the smectic in the (large) window of length scales between the
intrinsic orientational correlation length of the aerogel and the
distance between disorder-induced smectic dislocations.

The anharmonic terms included in Eq.\ref{H} are required by the
underlying global rotational invariance of the smectic phase\cite{GP},
hidden by the {\em spontaneous} choice of the layers to stack along
$\hat{\bf z}$-direction.

To compute self-averaging quantities, e.g., the disorder averaged free
energy, it is convenient (but not necessary) to employ the replica
``trick''\cite{Anderson} that relies on the identity $\overline{\log
Z}=\lim_{n\rightarrow 0}{\overline{Z^n}-1\over n}$. This allows us to
work with a translationally invariant field theory at the expense of
introducing $n$ replica fields (with the $n\rightarrow 0$ limit to be
taken at the end of the calculation). After replicating and
integrating over the disorder ${\bf h}({\bf r})$ utilizing
Eq.\ref{Delta}, we obtain:
\begin{eqnarray}
H[u_\alpha]&=&{1\over2}\int_{\bf r}
\sum_{\alpha=1}^n\bigg[{K}(\nabla_\perp^2
u_\alpha)^2+{B}(\partial_z u_\alpha- 
{1\over2}(\bbox{\nabla}_\perp u_\alpha)^2\big)^2\bigg]\nonumber\\
&-&{\Delta\over 2T}\int_r\sum_{\alpha,\beta=1}^n
\bbox{\nabla_\perp}u_\alpha\cdot\bbox{\nabla_\perp}u_\beta
\label{Hr}
\end{eqnarray}
from which the noninteracting propagator $G_{\alpha\beta}({\bf
q})\equiv V^{-1}\langle u_\alpha({\bf q})u_\beta(-{\bf q})\rangle_0$ can be
easily obtained,
\begin{equation}
G_{\alpha\beta}(q)=T G({\bf q})\delta_{\alpha\beta}+
\Delta q_\perp^2 G({\bf q})^2\;,
\label{G}
\end{equation}
with $G({\bf q})=1/(K q_\perp^4+B q_z^2)$. The fluctuations associated
with the disorder (the term in Eq.\ref{G} proportional to $\Delta$)
are much larger, as ${\bf q} \rightarrow {\bf 0}$, than those
associated with thermal fluctuations (the term proportional to $T$).

\begin{figure}[bth]
\centering
\setlength{\unitlength}{1mm}
\begin{picture}(18,18)(0,0)
\put(-52,-83){\begin{picture}(30,30)(0,0)
\includegraphics{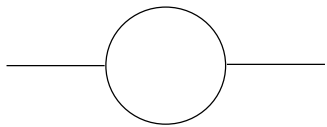}
\end{picture}}
\end{picture}
\caption{Feynman graph that renormalizes the elastic moduli $K$, $B$
and the disorder variance $\Delta$.}
\label{fig}
\end{figure}

We first attempt to assess the effects of the anharmonicities, disorder
and thermal fluctuations by performing a simple perturbation expansion
in the nonlinearities of $H[u_\alpha]$. The lowest order correction
$\delta B$ to the bare elastic compressional modulus $B$ comes from a
part of the diagram in Fig.\ref{fig}.
A simple analysis gives
\begin{eqnarray}
\delta B&=&-{B^2\over2}\int_{\bf q}^>\left[T G({\bf q})^2
+2\Delta q_\perp^2 G({\bf q})^3\right]q_\perp^4\;,\label{deltaBa}\\
&\approx&-B^2\Delta\int_{-\infty}^{\infty}{d q_z\over2\pi}
\int^>{d^{d-1}q_\perp\over(2\pi)^{d-1}}
{q_\perp^6\over(K q_\perp^4+B q_z^2)^3}\;,\label{deltaBb}\\
&\approx& -{3\over16} {C_{d-1}\over 5-d}\,
\Delta\left({B\over K^5}\right)^{1/2} L^{5-d}\;,\label{deltaBc}
\end{eqnarray}
where the constant $C_d=2\pi^{d/2}/\left((2\pi)^d\Gamma(d/2)\right)$,
we have dropped the thermal contribution that is subdominant to the
disorder part, shown only the dominant contribution for $d<5$, and
introduced a long wavelength cutoff $L$ (defined by restricting the
wavevector integral to $q_\perp>1/L$). The divergence of this
correction as $L\rightarrow\infty$ signals the breakdown of
conventional harmonic elastic theory on length scales longer than
$\xi_{NL}^\perp$, which we define as the value of $L$ at which
$|\delta B(\xi_{NL}^\perp)|=B$. This definition gives:
\begin{equation}
\xi_{NL}^\perp=\left({16(5-d)K^{5/2}\over3
C_{d-1}B^{1/2}\Delta}\right)^{1/(5-d)}\;,\label{xiNL}
\end{equation}
which for physical 3d smectics is given by $\xi_{NL}^\perp=\left(64\pi
K^{5/2}/3B^{1/2}\Delta\right)^{1/2}$. We can also obtain a non-linear
crossover length in the $z$-direction
$\xi_{NL}^z=(\xi_{NL}^\perp)^2/\lambda\sim K^2/\Delta$, where
$\lambda\equiv(K/B)^{1/2}$, by imposing the infra-red cutoff in the
$z$-direction.

To understand the physics beyond this crossover scale, i.e. to make
sense of the apparent infra-red divergences found in Eq.\ref{deltaBc},
we employ the standard momentum shell renormalization group (RG)
transformation. We separate the displacement field into high and low
wavevector components: $u_\alpha({\bf r})=u^<_\alpha({\bf
r})+u^>_\alpha({\bf r})$, where $u^>_\alpha$ has support in the
wavevector range $\Lambda e^{-\ell}<q_\perp<\Lambda$ and $\Lambda$ is
an ultraviolet cutoff of order $1/\xi_{NL}^\perp$, integrate out the
high wavevector part $u^>_\alpha({\bf r})$, and rescale the lengths
and long wavelength part of the fields with
$r_\perp=r_\perp'e^{\ell}$, $z=z'e^{\omega\ell}$ and $u^<_\alpha({\bf
r})= e^{\chi\ell}u_\alpha'({\bf r'})$, so as to restore the uv cutoff
back to $\Lambda$. The underlying rotational invariance insures that
the graphical corrections preserve the rotationally invariant operator
$\big(\partial_z u_\alpha - {1\over2}(\bbox{\nabla}_\perp
u_\alpha)^2\big)$, renormalizing it as a whole. It is therefore
convenient (but not necessary) to choose the dimensional rescaling
that also preserves this operator; the appropriate choice is
$\chi=2-\omega$. This rescaling then leads to the zeroth order RG
flows of the effective couplings $K(\ell)=K e^{(d-1-\omega)\ell}$,
$B(\ell)=B e^{(d+3-3\omega)\ell}$, and $(\Delta/T)(\ell)=(\Delta/T)
e^{(d+1-\omega)\ell}$. From these dimensional couplings one can
construct two dimensionless couplings $\tilde{g}_1\equiv(B/K^3)^{1/2}$
and $\tilde{g}_2\equiv\Delta(B/K^5)^{1/2}$, whose flow is given by
$\tilde{g_1}(\ell)=\tilde{g_1}e^{(3-d)\ell}$ and
$\tilde{g_2}(\ell)=\tilde{g_2} e^{(5-d)\ell}$. $\tilde{g_1}$ is just
the coupling that becomes relevant in $d<3$ and was discovered in
Ref.\cite{GP} to lead to anomalous elasticity in pure smectics. It is,
however, only marginally irrelevant in $d=3$ and therefore only leads
to a weak anomalous elasticity in physical 3d smectics. In contrast,
the upper critical dimension $d_{uc}$ below which $\tilde{g_2}$
becomes relevant is $d_{uc}=5$, and leads to much stronger anomalous
elasticity, that should be experimentally observable in disordered 3d
smectics. These observations imply that temperature is a strongly
irrelevant variable near the disorder dominated fixed point. We will
therefore set $T=0$ in all subsequent calculations.

The integration over the high wavevector components of $u_\alpha$ can
only be accomplished perturbatively in nonlinearities of $H[u]$. This
perturbation theory can be represented graphically; the graph giving
the leading order corrections $\delta B$, $\delta K$ and
$\delta\Delta$ (with the part diagonal in the replica indices
$\alpha,\beta$ renormalizing $K$ and $B$, and the part independent of
$\alpha,\beta$ correcting $\Delta$) is shown in
Fig.\ref{fig}.\cite{comment} Evaluating it, and performing the
rescalings described above, we obtain the following RG flow equations
\begin{eqnarray}
\frac{d B(\ell)}{d\ell}&=&(d+3-3\omega-{3\over16}g_2)B\;,\label{B}\\
\frac{d K(\ell)}{d\ell}&=&(d-1-\omega+{1\over32}g_2)K\;,\label{K}\\
\frac{d(\Delta/T)(\ell)}{d\ell}&=&(d+1-\omega+{1\over64}g_2)(\Delta/T)\;,
\label{DeltaT}
\end{eqnarray}
where we have defined a dimensionless coupling constant
$g_2\equiv\Delta(B/K^5)^{1/2}C_{d-1}\Lambda^{d-5}$, which obeys
\begin{equation}
\frac{d g_2(\ell)}{d\ell}=\epsilon g_2 - {5\over32} g_2^2\;,
\label{g2}
\end{equation}
with $\epsilon\equiv5-d$.  As required the flow of $g_2$ is
independent of the arbitrary choice of the anisotropy rescaling
exponent $\omega$. The RG flow Eq.\ref{g2} shows that the Gaussian
$g_2^*=0$ fixed point becomes unstable for $d<5$, and the low
temperature phase is controlled by a stable, nontrivial, glassy $T=0$
fixed point at $g_2^*=32\epsilon/5$.

The existence of this nontrivial fixed point leads to the anomalous
elasticity, which we can calculate using the following matching
approach. For this purpose it is convenient to use our RG results to
evaluate the connected disordered averaged 2-point $u({\bf k})$
correlation function $G({\bf k})\propto\overline{\langle|u({\bf
k})|^2\rangle}- \overline{\langle u({\bf k})\rangle\langle u(-{\bf
k})\rangle}$. The power of the renormalization group is that it
establishes a connection between a correlation function at a small
wavevector (which is impossible to calculate in perturbation theory
due to the infra-red divergences) to the same correlation function at
large wavevectors, which can be easily calculated in a controlled
perturbation theory. This relation for $G({\bf k})$ is
\begin{eqnarray}
G({\bf k}_\perp, k_z, K, B, g_2)&=&\label{relationG}\\
&&\hspace{-.5in}e^{(3+d-\omega)\ell}
G({\bf k}_\perp e^\ell, k_z e^{\omega\ell}, K(\ell), B(\ell),
g_2(\ell))\;,\nonumber
\end{eqnarray}
where the prefactor on the right hand side comes from the dimensional
rescaling using the exact Ward identity $\chi=2-\omega$, and we have
traded in the disorder variable $\Delta$ for the dimensionless
coupling $g_2$. To establish the anomalous behavior of $K$, we look at
$k_z=0$. We then choose the rescaling variable $\ell^*$ such that
$k_\perp e^{\ell^*}=\Lambda$. We also choose $k_\perp$ sufficiently
small such that $g_2(\ell^*)$ has reached our nontrivial fixed point
$g_2^*$. Eliminating $\ell^*$ in favor of $k_\perp$, we then obtain
\begin{eqnarray}
G({\bf k}_\perp, 0, K, B, g_2)&=&\label{relationG2}\\
&&\hspace{-.5in}\left(\Lambda\over k_\perp\right)^{3+d-\omega}
G\left(\Lambda, 0,K(\ell^*),B(\ell^*),g_2^*\right)\;,\nonumber
\end{eqnarray}
Since the right hand side is evaluated at 
the Brillouin zone boundary, it can be calculated perturbatively in
the fixed point coupling $g_2^*$. To lowest order we
obtain
\begin{eqnarray}
G({\bf k}_\perp, 0, K, B, g_2)&\approx&
{(\Lambda/k_\perp)^{3+d-\omega}\over\Lambda^4
K\left(\Lambda/k_\perp\right)^{(d-1-\omega+1g_2^*/32)}}
\;,\label{relationG3}\\
&\equiv&{1\over K(k_\perp) k_\perp^4}
\end{eqnarray}
where we integrated Eq.\ref{K} to obtain $K(\ell^*)$, and defined the
anomalous tilt modulus which diverges at long length scales
\begin{equation}
K(k_\perp)=K\left(k_\perp/\Lambda\right)^{-\eta_K}\;,\label{Kanom}
\end{equation}
with an anomalous exponent
\begin{eqnarray}
\eta_K&=&{1\over32}g_2^*={1\over 5}\;\epsilon\;,\label{etaK2}\\
&=&{2\over5}\;,\;\;\;\mbox{for}\;\; d=3\;.\label{etaK3}
\end{eqnarray}
%

%Similar calculations for $B$ lead to 
%%
%\begin{equation}
%G(0, k_z, K, B, g_2)\equiv
%{1\over B(k_z) k_z^2}\;,
%\end{equation}
%%
%with the anomalous vanishing compression modulus
%%
%\begin{equation}
%B(k_z)=B\left(k_z/\Lambda_z\right)^{\eta_B/\zeta}\;,\label{Banom}
%\end{equation}
%%
%where the cutoff $\Lambda_z\equiv\Lambda^2 (K/B)^{1/2}$, the
%anomalous exponent 
%%
%\begin{eqnarray}
%\eta_B&=&{3\over16}g_2^*={6\over5}\;\epsilon\;,\label{etaB2}\\
%&=&{12\over5}\;,\;\;\;\mbox{for}\; d=3\;.\label{etaB3}
%\end{eqnarray}
%%
%and $\zeta\equiv 2-(\eta_B+\eta_K)/2$ is the anisotropy exponent, which
%would be 2 in the absence of anharmonic effects.  Similarly, the
%disorder variance is length scale dependent, diverging at long scales
%%
%\begin{equation}
%\Delta(k_\perp)\sim \Delta k_\perp^{-\eta_\Delta}\;,\label{Danom}
%\end{equation}
%%
%with the anomalous exponent 
%%
%\begin{eqnarray}
%\eta_\Delta&=&{1\over64}g_2^*={1\over10}\;\epsilon\;,\label{etaD2}\\
%&=&{1\over5}\;,\;\;\;\mbox{for}\; d=3\;.\label{etaD3}
%\end{eqnarray}
%%

Similar calculations for the other coupling constants and other
directions of $\bf k$ show that, in general, 
\begin{eqnarray}
K({\bf k})&=&K\left(k_\perp\xi_{NL}^\perp\right)^{-\eta_K}
f_K(k_z\xi_{NL}^{z}/(k_\perp\xi_{NL}^\perp)^\zeta)\;,\label{Kg}\\
B({\bf k})&=&B\left(k_\perp\xi_{NL}^\perp\right)^{\eta_B}
f_B(k_z\xi_{NL}^{z}/(k_\perp\xi_{NL}^\perp)^\zeta)\;,\label{Bg}\\
\Delta({\bf k})&=&\Delta\left(k_\perp\xi_{NL}^\perp\right)^{-\eta_\Delta}
f_\Delta(k_z\xi_{NL}^{z}/(k_\perp\xi_{NL}^\perp)^\zeta)\;,\label{Deltag}
\end{eqnarray}
with the anisotropy exponent $\zeta\equiv 2-(\eta_B+\eta_K)/2$, which
would be $2$ in the absence of anharmonic effects,
$\eta_B=3g_2^*/16=6\epsilon/5=12/5$ in $d=3$, and
$\eta_\Delta=g_2^*/64=\epsilon/10=1/5$ in $d=3$.

Of course, we do not completely trust the extrapolation of these small
$\epsilon$ results down to $\epsilon=2$ ($d=3$). However, since by
definition $d g_2/d\ell = 0$ at the nontrivial fixed point, this
condition implies an {\em exact} relation between the anomalous
exponents,
\begin{equation}
5-d + \eta_\Delta = {\eta_B\over 2} + {5\over 2}\eta_K\;,
\label{WI}
\end{equation}
which is obviously satisfied by the anomalous exponents, 
%Eqs.\ref{etaK2},\ref{etaB2},\ref{etaD2}, 
computed here to first order in $\epsilon$. This Ward identity between
the anomalous exponents can be equally easily obtained from a
self-consistent integral equation for the $u-u$ correlations
functions, using renormalized wavevector dependent elastic moduli and
disorder variance.\cite{RTaerogel}

%In general the anomalous moduli and disorder variance can be written in
%the following scaling form:
%%
%\begin{eqnarray}
%K({\bf k})&=&K\left(k_\perp\xi_{NL}^\perp\right)^{-\eta_K}
%f_K(k_z\xi_{NL}^{z}/(k_\perp\xi_{NL}^\perp)^\zeta)\;,\label{Kg}\\
%B({\bf k})&=&B\left(k_\perp\xi_{NL}^\perp\right)^{\eta_B}
%f_B(k_z\xi_{NL}^{z}/(k_\perp\xi_{NL}^\perp)^\zeta)\;,\label{Bg}\\
%\Delta({\bf k})&=&\Delta\left(k_\perp\xi_{NL}^\perp\right)^{-\eta_\Delta}
%f_\Delta(k_z\xi_{NL}^{z}/(k_\perp\xi_{NL}^\perp)^\zeta)\;.\label{Deltag}
%\end{eqnarray}
%%

At length scales beyond $\xi_{NL}^\perp$ and $\xi_{NL}^z$, the
elasticity and fluctuations of the disordered smectic are controlled
by our new glassy fixed point. One of the important consequences can
be seen in the layer fluctuations that can be observed in X-ray
scattering experiments. For instance, layer displacement fluctuations
along $z$ are described by 
\begin{eqnarray}
C(z)&\equiv&\overline{\langle\left(u(0_\perp,z)-
u(0_\perp,0)\right)^2\rangle}\;,\nonumber\\
&=&\int{d^{d}k\over(2\pi)^{d}} {2(1-\cos(k_z z))\Delta({\bf k})
k_\perp^2\over\big(K({\bf k}) k_\perp^4+B({\bf k}) k_z^2\big)^2}
\;,\label{Cz}
\end{eqnarray}
One can then naturally define the X-ray translational correlation
length $\xi_z^{X}$, as the length along $z$ at which
$C(z=\xi_z^{X})\equiv a^2$, where $a$ is the smectic layer spacing. A
simple calculation, using Eqs.\ref{Kg}-\ref{Deltag} leads in 3d to
\begin{equation}
\xi_z^{X}=\left(a/\lambda\right)^{2/\gamma}K^2/\Delta=
\left(a/\lambda\right)^{2/\gamma}\xi^z_{NL}\;,\label{xi_z}
\end{equation}
where $\gamma\equiv(\eta_B+\eta_K)/\zeta$. Note that this X-ray
correlation length is finite even as $T\rightarrow 0$. This result is
consistent with the experimental observation\cite{Clark} that the
X-ray correlation length for smectics in aerogel saturates at some
finite value at low temperatures. Note also that this length should be
different for different smectics in the same aerogel, since $ B, K$,
and $\Delta$ will change from smectic to smectic. Since we expect
$\Delta \propto\rho_A$, the aerogel density\cite{RTaerogel}, the
aerogel density dependence of $\xi_z^X$ could test the prediction of
Eq.\ref{xi_z}. Likewise, the {\em temperature} dependence of $\xi_z^X$
could be used to determine $\gamma$, since the {\em bulk} $K(T)$ and
$B(T)$ that implicitly appear in Eq.\ref{xi_z} have temperature
dependence that can be extracted from measurements on bulk materials.

Note also that this correlation length is longer than the non-linear
crossover length for $\lambda<a$ (i.e., for large $B$). For
$\lambda>>a$ (small $B$), $C(z)$ reaches $a^2$ before $z$ reaches
$\xi_{NL}^z$, and hence anharmonic effects are unimportant. In this
case, the correlation length $\xi_z^X$ can be determined in a harmonic
theory (which amounts to evaluating the integral in Eq.\ref{Cz} with
$K({\bf k})$, $B({\bf k})$, and $\Delta({\bf k})$ replaced by their
constant (bare) values $K$, $B$, and $\Delta$). This gives
$\xi_z^X=a^2 B K/\Delta=(a/\lambda)^2\xi^z_{NL}$, which is,
reassuringly, much less than $\xi^z_{NL}$ in the limit $a<<\lambda$ in
which we have asserted it applies.

Although our entire discussion so far has focussed only on the effects
of {\it orientational} disorder, we have shown\cite{RTaerogel} that
{\it translational} disorder (i.e., random pinning of the positions of
the layers, is {\it less} important, at long wavelengths (in $d<5$),
than the orientational disorder. Thus, the results described herein
are directly applicable to real smectics, where both kinds of disorder
are present.

We have also ignored dislocations so far. Further detailed
analysis\cite{RTaerogel} shows that in a harmonic theory dislocations
unbind for arbitrarily weak disorder.  However, their effects are only
felt on length scales greater than $\xi_{disl.}=
(a^2/\lambda)e^{t_0\lambda/\Delta}$, where $t_0$ is a parameter of
order $B K$, which is much longer, in the weak disorder
($\Delta\rightarrow0$) limit, than the translational correlation
length $\xi_z^X$ found here.  Including anomalous elasticity, we
find\cite{RTaerogel} a shorter $\xi_{disl.}$ that nonetheless remains
much longer than $\xi_z^X$.  Hence our predictions for $\xi_z^X$ and
the anomalous elasticity studied here remain valid.

In summary, we have found that the addition of disorder to a smectic
liquid crystal phase leads to anomalous elasticity, that is much
stronger than in the pure system. This elasticity is controlled by a
zero temperature glassy fixed point, perturbative in
$\epsilon=5-d$. We expect that in contrast to marginally weak
anomalous elasticity of pure smectics, the strong anomaly in the
elasticity of disordered smectics, described here, should be readily
observable.

Note added: after this work was completely we learned of earlier
work\cite{IF} on the model Eq.\ref{Hr} in the context of spin
glasses. However, this earlier work misses the essential physics of
the nontrivial renormalization of disorder, and therefore makes {\em
quantitatively and qualitatively wrong} predictions of nearly all
results; e.g. the anomalous exponents, disorder correlations, scaling
exponent relations, etc., correct versions of which are given in our
Eqs.\ref{DeltaT}-\ref{WI}.

L.R. and J.T. acknowledge useful discussions with N. Clark and support
by the NSF through Grants DMR-9625111 and DMR-9634596, respectively.
\vspace{-.2in}

\end{multicols} 
\end{document}